\documentclass[11pt]{article}

\usepackage{epsfig}
\usepackage{graphicx}
\usepackage{amsmath}
\usepackage{verbatim}
\DeclareGraphicsExtensions{.eps}

\date{}

\begin{document}

\title{Evolutionary Dilemmas in a Social Network}
\author{%
L. Luthi\thanks{Information Systems Department, University of Lausanne, Switzerland} \and 
E. Pestelacci%
\addtocounter{footnote}{-1}%
\footnotemark
\thanks{Information Systems Department, University of Lausanne, Switzerland}
\and M. Tomassini%
\addtocounter{footnote}{-1}%
\footnotemark
\thanks{Information Systems Department, University of Lausanne, Switzerland} 
}
\maketitle

\begin{abstract} 
\noindent We simulate the prisoner's dilemma and hawk-dove games on a real social
acquaintance network. Using a discrete analogue of replicator dynamics, we show that
surprisingly high levels of cooperation can be achieved, contrary to what happens in
unstructured mixing populations. Moreover, we empirically show that cooperation in this
network is stable with respect to invasion by defectors.    
\end{abstract}

\section{Introduction}
\label{intro}

Some extremely simple games lead to puzzles and dilemmas that
have a deep social meaning. The \textit{Prisoner's Dilemma} (PD), a universal metaphor for the tension
that exists between social welfare and individual selfishness, is the most famous game of this type.
It stipulates that, in situations where individuals may either cooperate or behave selfishly and thus defect,
they will rationally choose the latter.
Unfortunately, cooperation would be the preferred outcome when global welfare is considered.
Game theory \cite{vega-redondo-03} is the discipline that deals with such situations of conflict where two 
or more individuals must make
decisions that will mutually influence each other. It takes a view of collective systems
in which global social outcomes emerge as a result of the interaction of the individual
decisions made by each agent.
Another well known simplified model of many common important socio-economic situations is the \textit{Hawk-Dove} (HD) game. 
According to game theory, cooperative attitude
should vanish in the PD, and should be limited to a given fraction in the 
HD. This is also the case when large populations of individuals play the game
pairwise in a random manner and anonymously, as prescribed by evolutionary game theory \cite{weibull95}.
However, in controlled experiments it has been observed that cooperation actually emerges 
when the game is played by humans and in many other cases \cite{axe84,sally-95}.
A number of mechanisms have been invoked to explain the emergence of cooperative behavior:
repeated interaction, reputation, and belonging to a recognizable group have often been mentioned \cite{axe84}.
However, the work of Nowak and May \cite{nowakmay92} showed that simply arranging the players  in a spatial structure and allowing them to only interact with neighbors is sufficient to sustain a certain amount of cooperation even when the game is played anonymously and without repetition.
Nowak and May's study and much of the following work was based on regular structures such as two-dimensional grids (see also \cite{hauer-doeb-2004} for the HD case).
However, while two-dimensional grids may be realistic for ecological and some biological
applications, they are inadequate for modeling human networks of interactions as it has now become clear that many actual networks
have a structure that is neither regular nor random but rather of the \textit{small-world} type.
Roughly speaking, small-world networks are graphs in which any node is relatively close to any other node.
In this sense, they are similar to random graphs but unlike regular lattices.
However, in contrast with random graphs, they also have a certain amount of local structure,
as measured, for instance, by a quantity called the \textit{clustering coefficient} which
essentially represents the probability that two neighbors of a given node are themselves connected (see e.g. \cite{newman-03}). 
Thus, most real conflicting situations in economy and sociology are not well 
described neither by a fixed
geographical position of the players in a regular lattice nor by a mixing population, and it becomes relevant to study these dilemmas on other, more faithful social
structures. Some previous work has been done in this direction. We mention
Santos and Pacheco's work on scale-free networks
 \cite{santos-pach-05,santos-pach-06} and work on Watts--Strogatz
small-world graphs \cite{social-pd-kup-01,tom-luth-giac-06,watts99}. However, these
network types, although they have the right global ``statistical'' properties, are only an approximation of the actual topological properties of measured networks of interactions. 
In the present work we introduce a more socially relevant network and we emphasize the relationships between community structure and cooperation.
A recent work close to the present one in spirit is Holme et al. \cite{trusina-karate-club}.
However, the authors of \cite{trusina-karate-club}
only study the PD on a much smaller social network using a different, noisy,
strategy update rule, while we employ the more standard replicator dynamics on a larger
social network and also study the HD game.

The remainder of this paper is organized as follows. We first give a brief background on the
PD and HD. We then describe the main features of
social networks and we present an evolutionary game
model on a real collaboration network. We finally present  and discuss results of numerical simulations of the
model in terms of cooperation, community structure and stability.

\section{The Model}
\label{sect:model}

\subsection{Social Dilemmas}
\label{sd}

We first recall a few elementary notions on the PD and the HD games.
These are two-person, symmetric games in which each player has two possible strategies:
cooperate (C) or defect (D). In strategic form, these games have the payoff
bi-matrix shown in table \ref{matr}.
In this matrix, R stands for the \textit{reward}
the two players receive if they
both cooperate, P is the \textit{punishment} for bilateral defection, and T  is the
\textit{temptation}, i.e. the payoff that a player receives if it defects, while the
other cooperates. In this latter case, the cooperator gets the \textit{sucker's} payoff S.
For the PD, the payoff values are ordered numerically in the following way: $T > R > P > S$, 
while in the HD game $T > R > S > P$. Defection is always the best rational individual choice in the PD -- (D,D) is the unique Nash equilibrium and also an evolutionary stable strategy (ESS).
Mutual cooperation  would be preferable but it is a strongly dominated strategy. Thus the
dilemma is caused by the ``selfishness'' of the actors.
\begin{table}[!ht]
\begin{center}
{\normalsize
\begin{tabular}{c|cc}
 & C & D\\
\hline
C & (R,R) & (S,T)\\
D & (T,S) & (P,P)
\end{tabular}
}
\end{center}
\caption{Payoff matrix for $2 \times 2$ symmetric games.}\label{matr}
\end{table}
\enlargethispage{\baselineskip}
In the HD, when both players defect they each get the lowest payoff;
(C,D) and (D,C) are Nash equilibria of the game in pure strategies, and there is
a third equilibrium in mixed strategies where strategy D is played
with probability $1/(2\beta-1)$, and strategy C with probability $1 - 1/(2\beta-1)$, where
$\beta$ is another name for the temptation $T$.
The dilemma in this game is caused by ``greed'', i.e. players have a strong incentive
to ``bully'' their opponent by playing D, which is harmful for both if the outcome is (D,D).

\subsection{Social Networks}
\label{sn}

In standard evolutionary game theory \cite{vega-redondo-03,weibull95}, these di\-lem\-mas have been classically
studied by modeling the behavior of a large population in which randomly paired individuals
play the game in an anonymous manner. Non-rational players are ``hard-wired'' to play a given
strategy, and those faring better than average increase their share in the population.
The fixed points of these \textit{replicator dynamics} are evolutionarily stable
strategies, i.e.\ they cannot be invaded by a mutant strategy \cite{weibull95}.

In terms of networks of interaction, the ``mixing'' structure of the population would be represented by a complete graph, i.e. any individual may interact with
any other player. The advantage of the
mixing model is that it admits an approach by \textit{mean-field} methods,
which treat the system as being homogeneous, ignoring space dependences and correlations
\cite{weibull95}.
However, we know today that real social networks do not have this random structure.
Instead, they are of finite size, have heterogeneous connectivity, are often small worlds, in the sense that any
individual is only a few steps away from any other, and individuals cluster together in
communities \cite{am-scala-etc-2000,newman-03,watts99}. Therefore,  evolutionary games should be studied on more general types of graphs, to understand the limitations of the theory, and to extend it as far as
possible to structures encountered in real-life.

As stated in the introduction, numerical simulations of evolutionary games have been performed 
for degree-homogeneous
and degree-heterogeneous model graphs such as Watts--Strogatz and scale-free networks
\cite{social-pd-kup-01,santos-pach-05,santos-pach-06,tom-luth-giac-06,watts99}.
Here we go a step further and make use of a real social collaboration network, the genetic
programming coauthorship network.
This 
network is a small world with a connected giant component of 942 scientists and it has recently been analyzed \cite{gp-graph-gpem}. It has clusters and communities and it should be representative
of other similar human acquaintance networks. Watts--Strogatz networks \cite{watts99} are only a mathematical
construction and do not properly represent social networks.  As for model scale-free graphs,
most social
networks studied to date are not of the pure scale-free type, and show a faster decay of the tail
of the degree distribution \cite{am-scala-etc-2000,newman-03}. Intuitively, there must be
a cutoff in the number of acquaintances a given agent can have, and in many cases also a typical number
of acquaintances, which gives a scale to the network. Besides, it has been observed that
social networks have a higher clustering than the typical values reached in scale-free
graphs, another manifestation of the complex neighborhood structure of
 the network. Furthermore, the appearance of communities -- sets of densely connected vertices with sparse connections between the sets --
 is yet another typical feature found in social structures.
 Communities can highly influence the way information is propagated throughout the network or opinion
 formation is processed.
 Finally, we should make it clear that 
social networks are dynamical, i.e.\ new nodes
may join the network forming new links, and old nodes may leave it as social actors
come and go.
As a first approximation here we model a static network, thus ignoring
fluctuations and non-equilibrium phenomena.

\subsection{Model Parameters}
\label{params}

\paragraph{\bf Population Structure.}
We consider a population $P$ of players of size $N$.
Each individual $i$ in the population $P$ is represented as a vertex $v_i$  of a graph $G(V,E)$,
with $v_i \in V, \; \forall i \in P$. An interaction between two players $i$ and
$j$ is represented by the undirected edge $e_{ij} \in E$.
The number of neighbors $N(i)$ of player $i$ is the degree $k_i$ of vertex $v_i$. The average
degree of the network will be called $\bar k$. The terms vertex, node, individual, or player
shall be used interchangeably in the sequel; likewise for edge, link, interaction, and
acquaintance.

\paragraph{\bf Strategy Update Rules.}
To update the strategies of the individuals given an initial strategy distribution in the
population, we use a discrete analogue of replicator dynamics \cite{hauer-doeb-2004}.
Other socially meaningful strategy update policies could also be used, such as \textit{imitation of the best}
and \textit{proportional updating} \cite{hauer-doeb-2004,nowakmay92,tom-luth-giac-06}.
The replicator dynamics assumes that the share of the population playing a particular
strategy grows in proportion to how well this strategy is doing relative to the average
population payoff. 

Let $\Pi_x$ be a player $x$'s aggregated payoff and $k_x$ the number of neighbors $x$ has ($k_x$ can also be seen as the degree of the vertex representing $x$).
We define the replicator dynamics function $\phi(\Pi_j -\Pi_i)$ as being the probability function according
to which player $i$ adopts neighbor $j$'s strategy, namely
\begin{eqnarray}
\phi(\Pi_j -\Pi_i)  =
\begin{cases} \displaystyle\frac{\Pi_j - \Pi_i}{k_j\Pi_{M1} - k_i\Pi_{m1}} & \textrm{if $\Pi_j - \Pi_i > 0$}\\\\
0 & \textrm{otherwise,}
\end{cases}
\label{repl_dyn_eq2}
\end{eqnarray}
where $\Pi_{M1}$ (resp.\ $\Pi_{m1}$) is the maximum (resp.\ minimum)
payoff a player could get if it had only one neighbor.

\paragraph{\bf Payoff Calculation.}
There exist several possibilities for determining a player's utility or payoff.
One can 
define a player's payoff as being the sum (\textit{accumulated payoff}) of all
pair interactions with its nearest neighbors. Or it can be defined as the accumulated payoff
 divided by the number of interactions (\textit{average payoff}).
Accumulated and average payoff give the same results 
when considering degree-homogenous networks such as regular lattices.
Accumulated payoff seems more logical to use in degree-heterogeneous networks since it
reflects the very fact that players may have different numbers of neighbors in the network.
Average payoff, on the other hand, smoothes out the possible differences although it might
be justified in terms of the number of interactions that a player may sustain in a given time,
i.e.\ an individual with many connections is likely to interact less often with each of its neighbors than another that has a lower number of connections.
Also, if there is a cost to maintain a relationship, average payoff will roughly capture this fact, while it will be hidden if one uses accumulated payoff.
In this paper we use a form of accumulated payoff.

\paragraph{\bf Population Dynamics.}
Calling $C(t) = (s_1(t), s_2(t), \ldots , s_N(t))$ a \textit{configuration} of the population
strategies $s_i \in \{C,D\}$ at time
step $t$, 
the global \textit{synchronous} system dynamics leads to $C(t+1)$ by simultaneously updating
all the players' strategies according to the chosen rule.
Synchronous update, with its idealization of a global clock, is customary 
in spatial evolutionary games, and most results have been obtained using this model.
However, perfect synchronicity is only an abstraction as 
agents normally act at different and possibly uncorrelated moments. In spite of this, 
it has been shown that the
update mode does not fundamentally alter the results for replicator dynamics
\cite{hauer-doeb-2004}. We have also
checked that asynchronous update dynamics does not influence the system evolution in
a significant way and so, all results presented refer to synchronous systems.

\section{Simulation Results and Analysis}
\label{sym-res}

For each game, we can explore the entire game space by limiting our study to the variation
of only two parameters per game.
In the case of the PD, we  set $R=1$ and $S=0$,
and vary $1 \leq T \leq 2$ and $0 \leq P \leq 1$.
For the HD game, we set $R=1$ and $P=0$ and the two parameters are $1 \leq T \leq 2$ and $0 \leq S \leq 1$.
In the Prisoner's Dilemma case, $P$ is limited between $R=1$ and $S=0$
in order to respect the ordering of the payoffs ($T>R>P>S$) and $T$'s upper bound is equal to 2  due to the
$2R > T+S$ constraint.
Had we instead fixed $R=1$ and $P=0$, $T$ could be as big as desired,
provided $S \leq 0$ is small enough.
In the Hawk-Dove game, setting $R=1$ and $P=0$ determines the range of $S$ (since this time $T>R>S>P$)
and gives an upper bound of 2 for $T$, again due to the $2R > T+S$ constraint.
Note however, that the only valid value pairs of $(T,S)$ are those that satisfy the latter constraint.

The network is randomly initialized with 50\% cooperators and 50\% defectors.
In all cases, the parameters are varied between their two bounds by steps of 0.1.
For each set of values, we carry out 50 runs of 16000 time steps each.
Cooperation level is averaged over the last 1000 time steps, well after the transient equilibration period.

\subsection{Evolution of Cooperation}
\label{coop}

\begin{figure} [!ht]
\begin{center}
\begin{tabular}{cc}
\mbox{\includegraphics[width=5cm,height=4.5cm]{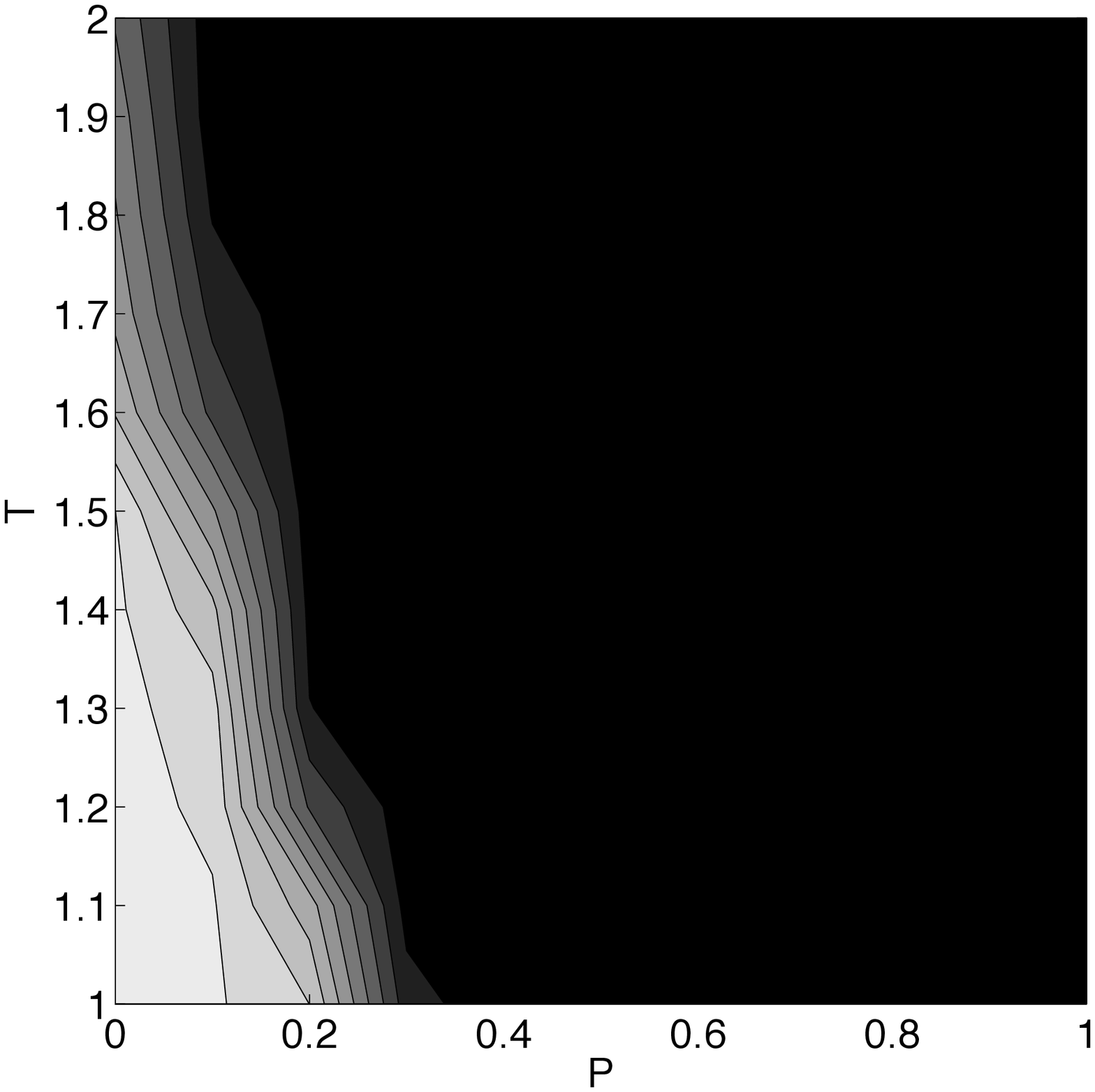}} \protect &
\hspace*{0.6cm}
\mbox{\includegraphics[width=5cm,height=4.5cm]{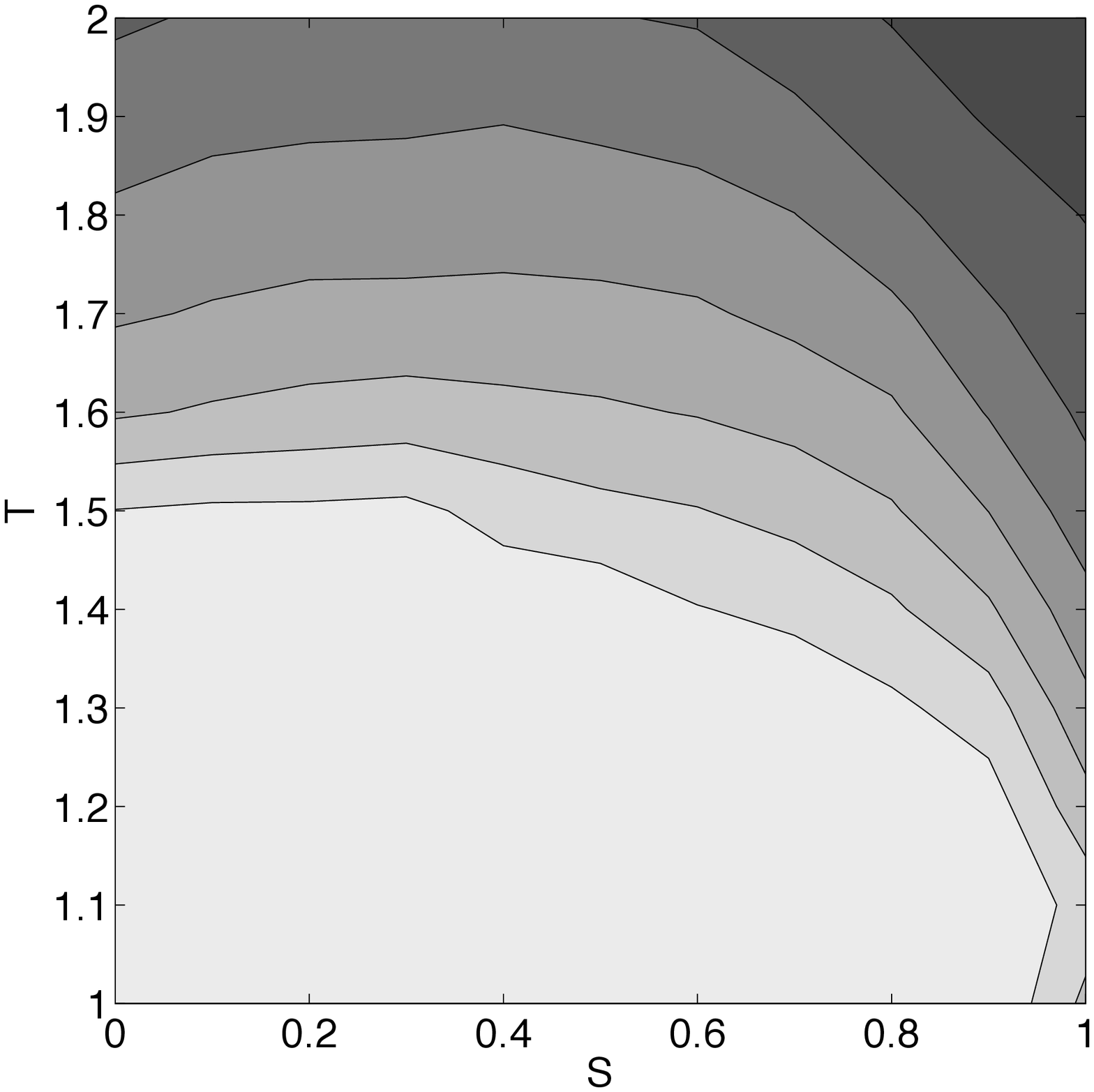}} \protect \\
 \hspace*{0.4cm}(a) & \hspace*{1cm} (b)
\end{tabular}
\caption{Level of cooperation at the end of the simulation. Left: PD; right: HD. For the HD, the meaningful phase space is the lower left triangle only. \label{coop-graphs}}
\end{center}
\end{figure}

In Figure \ref{coop-graphs} we report average cooperation levels
for both games  for systems having attained a
steady-state. As expected, the region in which cooperation is possible is much more restricted in the PD than for HD. Cooperation is more widespread for the HD, as mutual defection is the worst outcome in
this game. For the PD cooperation is sensitive to the ``punishment'' level P, for
a given T. Concerning the HD, one can see
that the S parameter has moderate influence on cooperation for a given T. We also notice that
the transition from cooperation to defection is much steeper in the PD than for the
HD.

\begin{figure} [!ht]
\begin{center}
\begin{tabular}{cc}
\hspace*{-0.3cm}\mbox{\includegraphics[width=6cm]{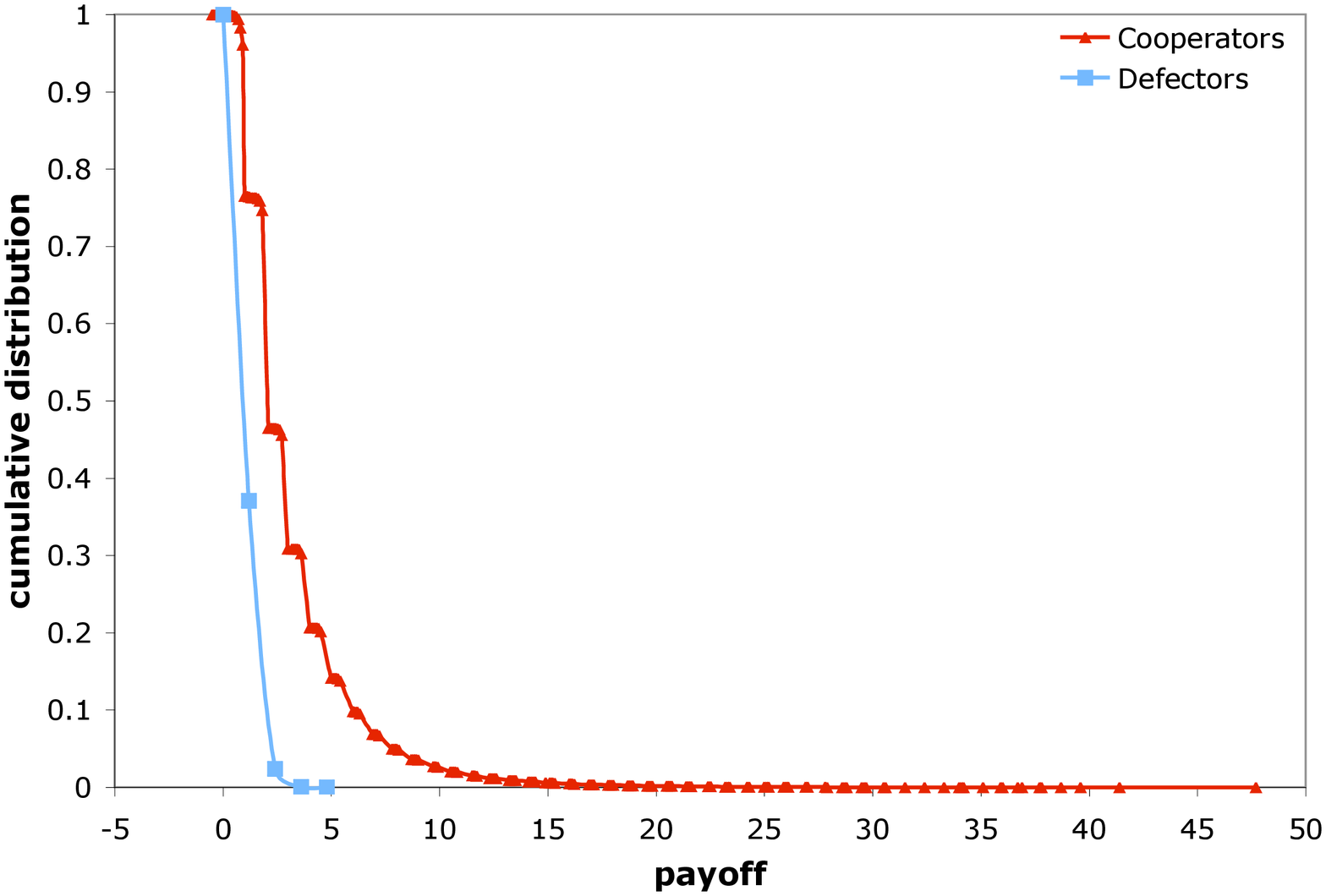}} \protect &
\mbox{\includegraphics[width=6cm]{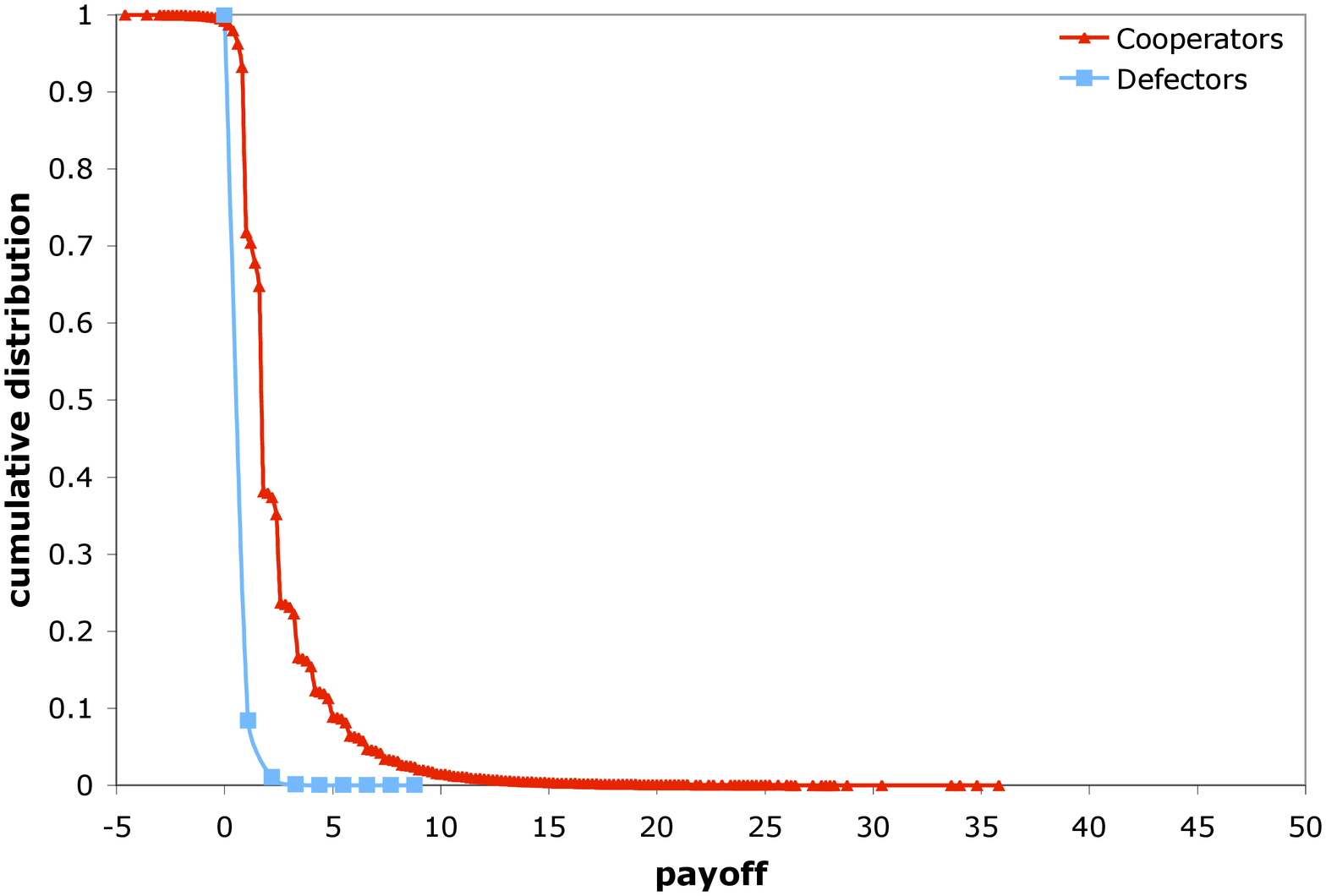}} \protect \\
\hspace*{-0.3cm}(a) & (b)
\end{tabular}
\caption{Cumulative wealth distribution in the PD game averaged over 50 runs for the social
network model; (a) $T=1.3, P=0.1$ yielding $\sim$73\% of cooperation. Average C-wealth=2.92, average D-wealth=0.47; (b) $T=1.3, P=0.2$ yielding $\sim$15\% of cooperation. Average C-wealth=2.30, average D-wealth=0.11.\label{wealth}}
\end{center}
\end{figure}

Another important global quantity  is the total payoff at the
end of the simulated games, also called the wealth. The cumulated wealth of defectors and
cooperators is plotted in Figure 
\ref{wealth} for the PD. This is done for $T=1.3$, for two values of the punishment P, giving rise to two different 
cooperation regimes: one in which cooperation prevails and a second one  where defection
predominates. We see that the cooperators' wealth is larger and has a broader distribution.
This hints at a clustering of cooperators, as this is the only way for them to increase their
payoff. We shall comment on this phenomenon below.

Social networks are characterized by the presence of communities, which can be seen as
sets of highly connected vertices having few connections with vertices belonging to other communities. Finding the communities in a given network is a difficult task for which
there exist several algorithms essentially based on clustering considerations. We have
used one of the algorithms proposed by Newman \cite{newman-2004-69}.

\begin{figure} [!ht]
\begin{center}
\mbox{\includegraphics[width=8cm]{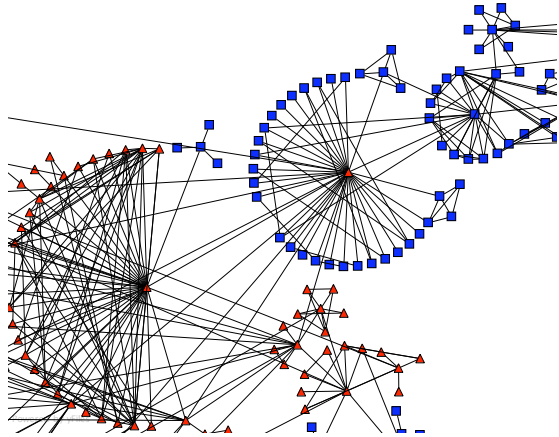}} \protect
\caption{Communities: cooperators are represented by triangles and defectors by squares.\label{communities}}
\end{center}
\end{figure}

For reasons of space, in the following we show results for the PD only.
In Figure \ref{communities} we depict a portion of the scientific coauthorship graph,
distinguishing between cooperators and defectors for the PD. We note that tightly-bound communities are mostly composed of players with the same strategy. 
Although we only
show a small portion of the whole network for reasons of clarity, we could have chosen many other places as the phenomenon is
widespread. 
Cooperators tend to ``protect'' themselves by having many links
toward other cooperators. On the other hand, a cooperator like the central one in the 
largest defecting  community will have
a tendency to become a defector since its neighbors are nearly all defectors; but when
its highly connected ``wealthy'' cooperator neighbor on the left of the figure is probabilistically selected to be
imitated, then it will certainly become a cooperator again. So, the rare cooperators that
are not tightly clustered with other cooperators will tend to oscillate between strategies.
The community structure of cooperators, together with the mutual payoff advantage of
cooperating, explains the previous observation, namely that the average
cooperators' wealth exceeds the average wealth of defectors. Strategy distribution in communities found for
the HD game is qualitatively similar; however, in the HD the two strategies
are slightly more intermingled, confirming analogous findings for grid-structured populations
\cite{hauer-doeb-2004}.

\enlargethispage{\baselineskip}

When cooperation prevails, we have found that cooperators tend to occupy the highest degree nodes.  When defectors predominate, the degree distributions of the two strategies tend to be closer,
although the cooperators still monopolize higher degree nodes. 
For HD the results are similar, namely, degree distribution for defectors fall off
more rapidly than those for cooperators. 

\subsection{Evolutionary Stability}
\label{Noise}

\begin{figure} [!ht]
\begin{center}
\begin{tabular}{cc}
\mbox{\includegraphics[width=5.8cm]{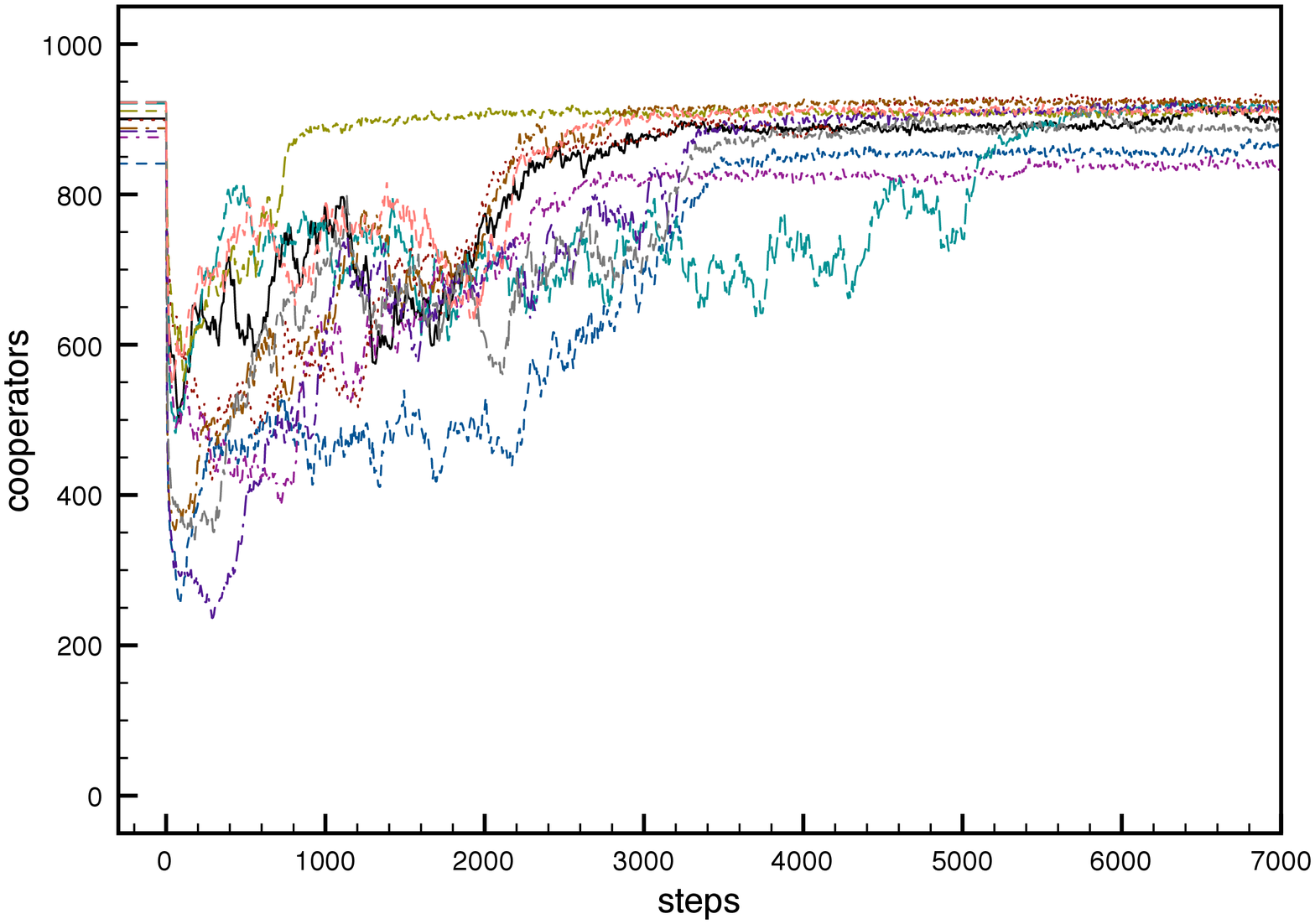}} \protect &
\hspace*{0.4cm}
\mbox{\includegraphics[width=5.8cm]{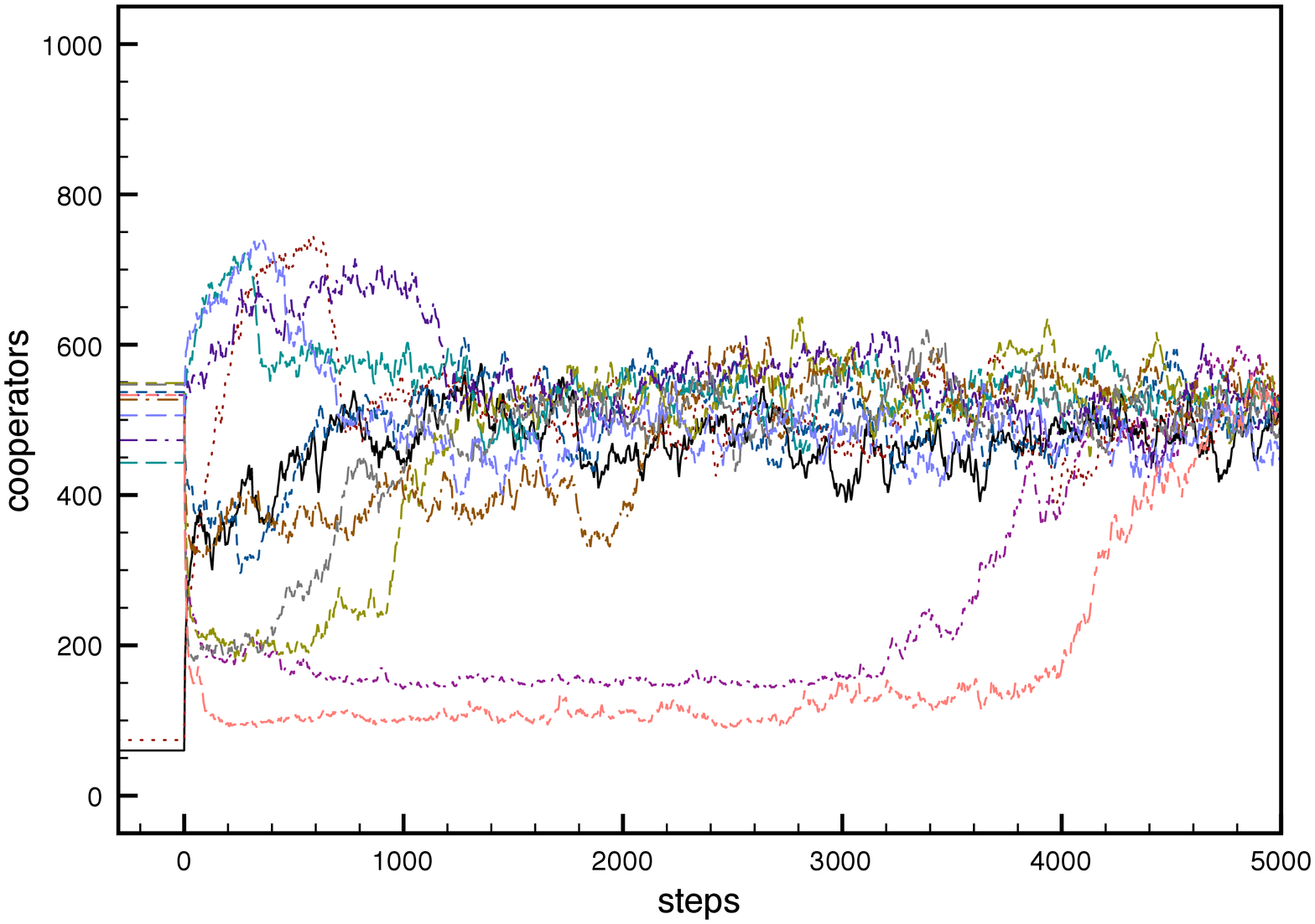}} \protect \\
 \vspace*{0.3cm}(a) & \hspace*{0.7cm} (b) \\
 \mbox{\includegraphics[width=5.8cm]{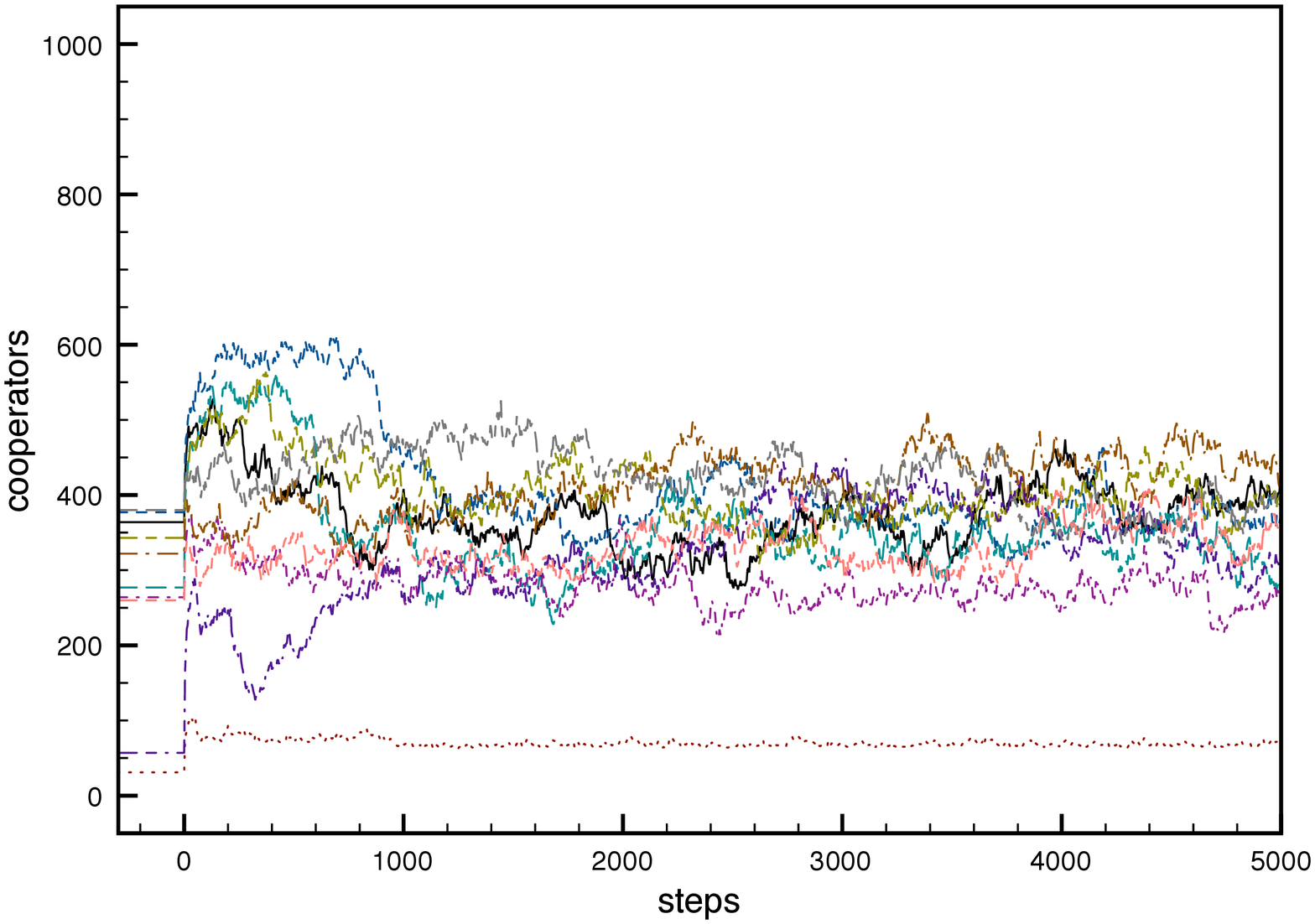}} \protect &
\hspace*{0.4cm}
\mbox{\includegraphics[width=5.8cm]{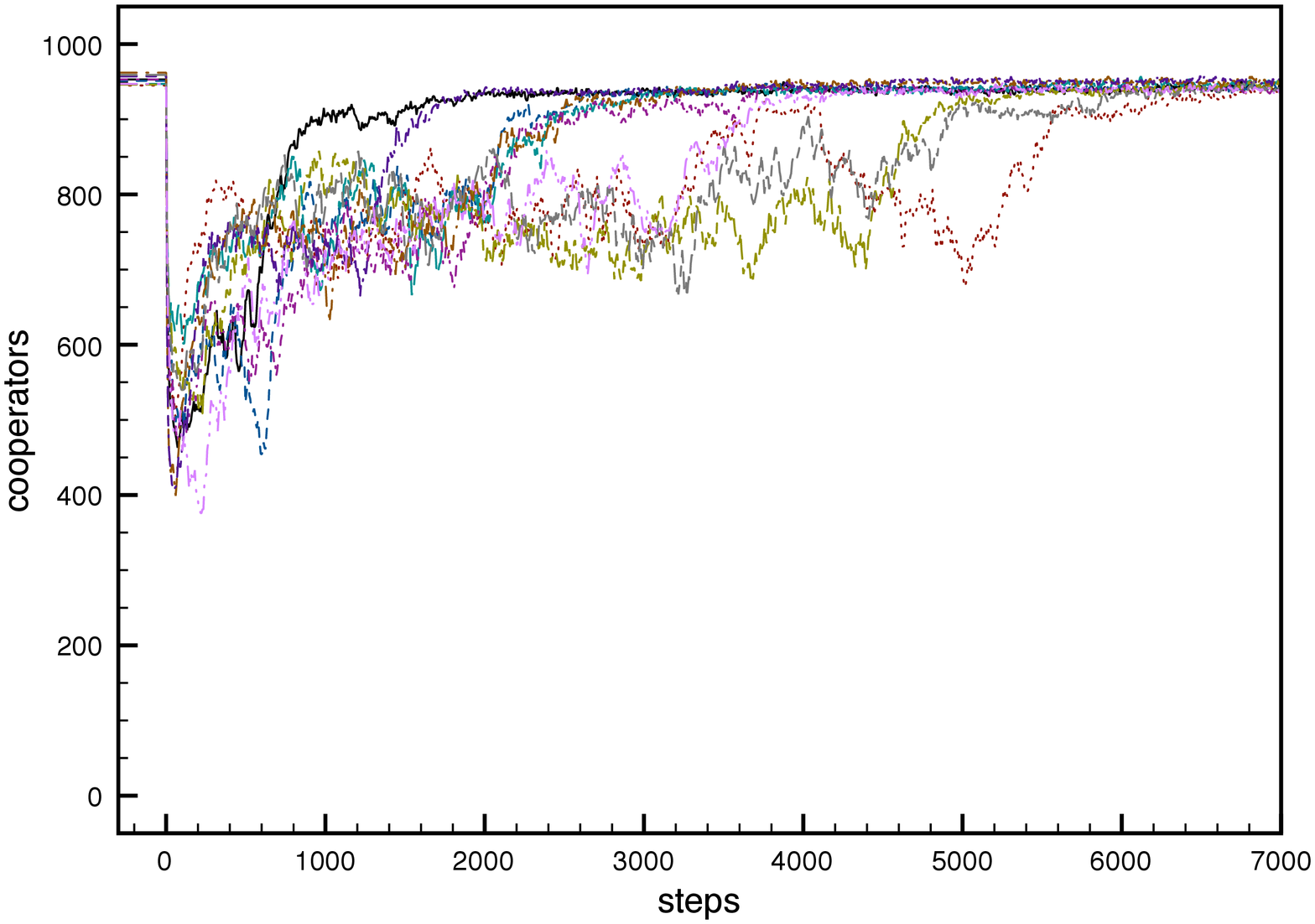}} \protect \\
 (c) & \hspace*{0.7cm} (d)
\end{tabular} 
\caption{Effect of noisy behavior on pseudo-equilibria states for the PD (a), (b), (c), and
HD (d) (see text for explanations). \label{noise}}
\end{center}
\end{figure}

No empirical investigation of an evolutionary games scenario would be complete without
examining its evolutionary stability \cite{weibull95}. Evolutionary stability can be
defined exactly for mixing populations but qualitatively it simply implies that a given population strategy cannot be invaded by an individual playing another strategy. For example,
a single defector in a mixing population of cooperators in the PD will lead to a total
extinction of the cooperators. Single
individual strategy mutations are interesting only in mixing populations or in networks of
the scale-free type \cite{santos-pach-06}. In our social network, the distribution of 
strategies that obtains when the steady-state is reached is left undisturbed by this kind
of event. However, given that the C and D strategies tend to cluster together, we have
applied a more radical type of perturbations to the system. After the pseudo-equilibrium
is reached, we choose a highly connected individual that plays the strategy of the majority
(suppose it is C) and we flip its strategy to D as well as the strategy of all its first
neighbors that are also cooperators.
In this case, evolutionary stability  requires that any such small group of individuals who try an
alternative strategy do worse than those that those who stick to the status quo.
Figure \ref{noise} shows the results using the above described perturbation. In each figure,
ten executions have been reported to give a feeling of the behavior (many more have been
run but in the case of noise, average values are irrelevant).
Figures \ref{noise} (a), (b), and (c) refer to the PD with $P=0.1$ and three different values
of $T$: $1.5, 1.7$, and $1.9$ respectively. Figure \ref{noise} (d) refers to the HD with $S=0.1,T=1.6$. The first constant part of the graphs up to abscissa 0 represent the pseudo-equilibrium values reached in each run after $10000$ steps, just before applying the
perturbation. For the PD, it can be seen that where cooperation is high (figure \ref{noise} (a)),
after a transient period in which there can be a significant loss of cooperation, all runs tend to
recover the original levels, albeit at different speeds. When cooperation is at intermediate
or lower levels (figures \ref{noise} (b) and (c)), the behavior is more oscillatory but populations
tend to recover the original pseudo-equilibrium levels of cooperation. Even populations that
had originally a significantly lower cooperation percentage can sometimes reach the
cooperation level of the majority of runs after the perturbation. For the
HD, populations easily recover from noisy behavior (see figure \ref{noise} (d)), as cooperation is more widespread in this game. Even for extreme values of $T$ the original cooperation level tends
to be recovered (not shown to save space but very similar to figure \ref{noise} (b)).

\section{Conclusions}
\label{concl}
Extending previous work on regular structures and on model scale-free and small-world
networks, in this paper we have empirically studied two fundamental social dilemmas on
a real acquaintance network. Although the graph studied is a single particular instance,
it possesses all the features that characterize actual social networks, such as high clustering
and communities. We find that this kind of topology allows cooperation to be reached
and maintained, for a large portion of the game parameter space for HD, and even in the more
difficult case of the PD. It was previously known that this is the case for lattice structures
and, most notably, for scale-free graphs. However, these structures are not good representations
of social ties and thus our result is encouraging from the social point of view. Importantly,
we have also shown that the quasi-equilibria reached by the dynamics are not ephemeral, unstable
states; on the contrary, they are very robust against perturbations represented by 
strategy flips of groups of agents. When reshuffled by the perturbation, even population configurations in which defection prevails either recover the previous level of cooperation or
increase it, which means that this kind of social graphs intrinsically favor cooperation through
clustering and tight communities.
In the future, we would like to generalize these results to classes of social network models.

\paragraph{\bf Acknowledgments.} Financial support for this research by the Swiss National Science Foundation under contract 200021-107419 is gratefully acknowledged.

\bibliographystyle{plain}

\end{document}